\documentclass[english,aps,prl,twocolumn]{revtex4-1}
\usepackage[T1]{fontenc}
\usepackage[latin9]{inputenc}
\usepackage{color}
\usepackage{amsmath}
\usepackage{amssymb}
\usepackage{graphicx}
\usepackage{esint}

\makeatletter
\@ifundefined{textcolor}{}
{%
 \definecolor{BLACK}{gray}{0}
 \definecolor{WHITE}{gray}{1}
 \definecolor{RED}{rgb}{1,0,0}
 \definecolor{GREEN}{rgb}{0,1,0}
 \definecolor{BLUE}{rgb}{0,0,1}
 \definecolor{CYAN}{cmyk}{1,0,0,0}
 \definecolor{MAGENTA}{cmyk}{0,1,0,0}
 \definecolor{YELLOW}{cmyk}{0,0,1,0}
 }

\makeatother

\usepackage{babel}
\begin{document}

\newcommand{\1}{{\bf \scriptstyle 1}\!\!{1}}
\newcommand{\unit}{\overleftrightarrow{{\bf \scriptstyle 1}\!\!{1}}}
\newcommand{\I}{{\rm i}}
\newcommand{\p}{\partial}
\newcommand{\D}{^{\dagger}}
\newcommand{\hbe}{\hat{\bf e}}
\newcommand{\bfa}{{\bf a}}
\newcommand{\bx}{{\bf x}}
\newcommand{\hbx}{\hat{\bf x}}
\newcommand{\by}{{\bf y}}
\newcommand{\hby}{\hat{\bf y}}
\newcommand{\br}{{\bf r}}
\newcommand{\hbr}{\hat{\bf r}}
\newcommand{\bj}{{\bf j}}
\newcommand{\bk}{{\bf k}}
\newcommand{\bn}{{\bf n}}
\newcommand{\bv}{{\bf v}}
\newcommand{\bp}{{\bf p}}
\newcommand{\tp}{\tilde{p}}
\newcommand{\tbp}{\tilde{\bf p}}
\newcommand{\bu}{{\bf u}}
\newcommand{\hbz}{\hat{\bf z}}
\newcommand{\bA}{{\bf A}}
\newcommand{\calA}{\mathcal{A}}
\newcommand{\calB}{\mathcal{B}}
\newcommand{\tC}{\tilde{C}}
\newcommand{\bD}{{\bf D}}
\newcommand{\bE}{{\bf E}}
\newcommand{\calF}{\mathcal{F}}
\newcommand{\bB}{{\bf B}}
\newcommand{\bG}{{\bf G}}
\newcommand{\calG}{\mathcal{G}}
\newcommand{\obG}{\overleftrightarrow{\bf G}}
\newcommand{\bJ}{{\bf J}}
\newcommand{\bK}{{\bf K}}
\newcommand{\bL}{{\bf L}}
\newcommand{\tL}{\tilde{L}}
\newcommand{\bP}{{\bf P}}
\newcommand{\calP}{\mathcal{P}}
\newcommand{\bQ}{{\bf Q}}
\newcommand{\bR}{{\bf R}}
\newcommand{\cR}{{\cal R}}
\newcommand{\bS}{{\bf S}}
\newcommand{\bH}{{\bf H}}
\newcommand{\balpha}{\mbox{\boldmath $\alpha$}}
\newcommand{\talpha}{\tilde{\alpha}}
\newcommand{\bsigma}{\mbox{\boldmath $\sigma$}}
\newcommand{\hbeta}{\hat{\mbox{\boldmath $\eta$}}}
\newcommand{\bSigma}{\mbox{\boldmath $\Sigma$}}
\newcommand{\bomega}{\mbox{\boldmath $\omega$}}
\newcommand{\bpi}{\mbox{\boldmath $\pi$}}
\newcommand{\bphi}{\mbox{\boldmath $\phi$}}
\newcommand{\hbphi}{\hat{\mbox{\boldmath $\phi$}}}
\newcommand{\btheta}{\mbox{\boldmath $\theta$}}
\newcommand{\hbtheta}{\hat{\mbox{\boldmath $\theta$}}}
\newcommand{\hbxi}{\hat{\mbox{\boldmath $\xi$}}}
\newcommand{\hbzeta}{\hat{\mbox{\boldmath $\zeta$}}}
\newcommand{\brho}{\mbox{\boldmath $\rho$}}
\newcommand{\bnabla}{\mbox{\boldmath $\nabla$}}
\newcommand{\bmu}{\mbox{\boldmath $\mu$}}
\newcommand{\bepsilon}{\mbox{\boldmath $\epsilon$}}

\newcommand{\iLambda}{{\it \Lambda}}
\newcommand{\cL}{{\cal L}}
\newcommand{\cH}{{\cal H}}
\newcommand{\cU}{{\cal U}}
\newcommand{\cT}{{\cal T}}

\newcommand{\be}{\begin{equation}}
\newcommand{\ee}{\end{equation}}
\newcommand{\bea}{\begin{eqnarray}}
\newcommand{\eea}{\end{eqnarray}}
\newcommand{\beqa}{\begin{eqnarray*}}
\newcommand{\eeqa}{\end{eqnarray*}}
\newcommand{\nn}{\nonumber}
\newcommand{\DD}{\displaystyle}

\newcommand{\ba}{\begin{array}{c}}
\newcommand{\baa}{\begin{array}{cc}}
\newcommand{\baaa}{\begin{array}{ccc}}
\newcommand{\baaaa}{\begin{array}{cccc}}
\newcommand{\ea}{\end{array}}

\newcommand{\bma}{\left[\begin{array}{c}}
\newcommand{\bmaa}{\left[\begin{array}{cc}}
\newcommand{\bmaaa}{\left[\begin{array}{ccc}}
\newcommand{\bmaaaa}{\left[\begin{array}{cccc}}
\newcommand{\ema}{\end{array}\right]}

\title{\textcolor{black}{Plasmonically enhanced  spectrally selective narrowband MWIR and LWIR light detection based on hybrid nanopatterned graphene and phase changing vanadium oxide heterostructure operating close to room temperature}}

\author{\textcolor{black}{Muhammad Waqas Shabbir$^{(1)}$}}

\author{\textcolor{black}{Sayan Chandra$^{(2)}$}}

\author{\textcolor{black}{Michael N. Leuenberger$^{(1,3)}$}}

\email{michael.leuenberger@ucf.edu}

\affiliation{$^{(1)}$ NanoScience Technology Center and Department of Physics, University of Central Florida, Orlando, FL 32826, USA. \\
$^{(2)}$ Department of Physics and Astronomy, Appalachian State University, Boone, NC 28608, USA. \\
$^{(3)}$ College of Optics and Photonics, University of Central Florida, Orlando, FL 32826, USA. }

\begin{abstract}
We present the model of an ultrasensitive mid-infrared (mid-IR) photodetector operating in the mid-wavelength infrared (MWIR) and long-wavelength infrared (LWIR) domains consisting of a hybrid heterostructure made of nanopatterned graphene (NPG) and vanadium dioxide (VO$_2$) which exhibits a large responsivity of  $R\sim 10^4$ V/W, a detectivity exceeding $D^*\sim 10^{10}$ J, and a sensitivity in terms of noise-equivalent power $\mathrm{NEP}\sim 100$ fW/$\sqrt{\rm Hz}$ close to room temperature by taking advantage of the phase change of a thin VO$_2$ film. Our proposed photodetector can reach an absorption of nearly 100\% in monolayer graphene due to localized surface plasmons (LSPs) around the patterned circular holes. The geometry of the nanopattern and an electrostatic gate potential can be used to tune the absorption peak in the mid-IR regime between 3 and 12 $\mu$m. After the photon absorption by the NPG sheet and the resulting phase change of VO$_2$ from insulating to metallic phase the applied bias voltage $V_b$ triggers a current through the VO$_2$ sheet, which can be detected electronically in about 1 ms, shorter than the detection times of current VO$_2$ bolometers. Our envisioned mid-IR photodetector reaches detectivities of cryogenically cooled HgCdTe photodetectors and sensitivities larger than VO$_2$ microbolometers while operating close to room temperature.

\textcolor{black}{KEYWORDS: Localized surface plasmons, graphene, absorbance, vanadium dioxide, photodetection, bolometer. }
\end{abstract}
\maketitle
Due to the low photon energy of IR radiation cryogenic cooling is required for highly sensitive photodetection based on low band gap materials like mercury-cadmium-telluride (HgCdTe). Various kinds of microbolometers primarily based on vanadium oxide (VOx) offer uncooled detection of IR radiation. However, microbolometers suffer from low sensitivity, slow response and tedious multi-step complex lithographic processes \cite{Rogalski2002}. 
Photodetection based on the bolometric effect takes advantage of the dependence of the resistivity on the temperature to detect incident light, typically in the infrared regime.
Vanadium oxide (VO$_2$) has become one of the standard materials for building microbolometers with broadband mid-IR photodetection because it features a reversible insulator-to-metal  phase transition (IMT) when heated above the phase transition temperature $T_c$, which is slightly above and close to room temperature \cite
{Rogalski2011}.
Bulk VO$_2$ undergoes a phase transition from an insulating state with monoclinic crystal structure below 68$^\circ$C to a metallic state with rutile crystal structure above $T_c=68^\circ$C (=341 K) \cite{Goodenou1971,Morin1959,Chandra2018}. This phase transition is fully reversible with a hysteresis loop, occurs on a subpicosecond timescale \cite{Appavoo2011,Dicken2009}, and can be initiated either thermally, electrically \cite{Stefanovich2000}, or optically \cite{Cavalleri2001}. While for bulk VO$_2$ optically induced IMT can only be achieved by pumping above the band gap of $E_g=670$ meV of bulk VO$_2$, it is possible to induce IMT in thin films at energies of 200 meV and above (corresponding to wavelengths of $\lambda=6.2$ $\mu$m and below)  due to electronic defects inside the band gap \cite{Rini2008}, which is the reason why VO$_2$ thin films can be used for IR detection in a wide IR range, in particular in the 3-5 $\mu$m range. Interestingly, when VO$_2$ is in the form of a thin film, its transition temperature $T_c$ depends strongly on the thickness d of the film, i.e. $T_c$ decreases from 65$^\circ$C (=338 K) for $d=25$ nm down to 52$^\circ$C (=325 K) for $d=3$ nm \cite{Xu2005}. While IR radiation with wavelengths above about $\lambda=1$ $\mu$m cannot detect the change in thickness of around 20 nm, it certainly distinguishes between the insulating and the metallic phase of VO$_2$. All these properties make VO$_2$ the ideal material for developing mid-IR photodetectors based on the IMT effect.

\begin{figure}[htb]
\begin{centering}
\includegraphics[width=7.5cm]{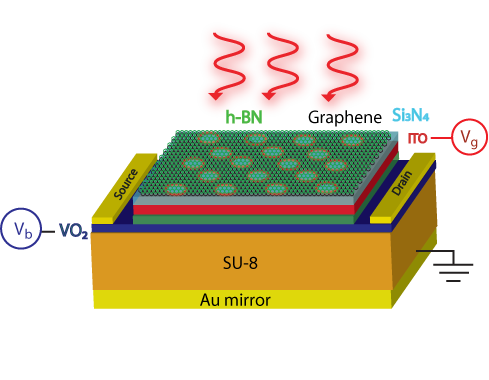}
\end{centering}
\caption{Schematic showing our proposed IMT-based mid-IR light photodetector consisting of hybrid NPG-VO$_2$.
The materials from top to bottom are:
1 single layer of hexagonal boron nitride (h-BN), for preventing oxidation of graphene at higher temperatures (*),
1 single layer of patterned graphene (*),
65 nm of Si$_3$N$_4$ (*),  for large n-doping and gating,
65 nm of ITO (*), metallic contact for gating, which is also transparent in mid-IR,
20 nm of h-BN (*), used for efficient heat transfer,
3 nm of VO$_2$ (*), contacted with source and drain Au leads,
$\lambda/4n_{\rm SU8}$ of SU8,\cite{Safaei2017} which is transparent in mid-IR, and
Au back mirror. $n_{\rm SU8}=1.56$ is the refractive index of SU8.
All the layers marked with an asterix (*) are patterned with the same hexagonal lattice of holes.
\label{fig:graphene_VO2_detector} }
\end{figure}

 However, photodetection of mid-IR light with wavelength above about 6 $\mu$m is inefficient with VO$_2$ bulk or thin films because of the relatively weak interaction between the incident photons and the optical phonons in VO$_2$. This interaction is so weak that the IMT cannot be achieved. That is why in the wavelength regime of 8 to 12 $\mu$m the photodetection is based purely on the bolometric effect in the semiconducting phase of VO$_2$ \cite{Barker1966,Chen2001}. 
 Here, by adding a single layer of nanopatterned graphene (NPG) on top of a layer of VO$_2$, we present the model of a photodetector that not only greatly enhances the absorption of mid-IR light energy in the longer wavelength regime from $\lambda=6$ $\mu$m and exceeding 12 $\mu$m but also narrows the absorption bandwidth to 0.1 $\mu$m within the mid-IR range of 3 to 12 $\mu$m, thereby enabling plasmonically enhanced spectrally selective absorption of mid-IR light for the IMT effect  in a heterostructure made of NPG and VO$_2$.
 
 Building upon the knowledge we acquired for developing spectrally selective photodetectors made of NPG that detect mid-IR light by means of the photothermoelectric effect \cite{Safaei2017,SafaeiACS,Safaei2019} and the knowledge we acquired for developing thermal emitters based on NPG \cite{Shabbir2020}, we develop here the model of a mid-IR microbolometer that consists of an hexagonal boron nitride (h-BN) coated NPG, silicon nitride (Si$_3$N$_4$), indium tin oxide (ITO), VO$_2$, polymer, and gold (Au) mirror, as shown in Fig.~\ref{fig:graphene_VO2_detector}. 
The main working principle of our envisioned mid-IR photodetector can be summarized as follows. After mid-IR photons get absorbed by the NPG sheet at a wavelength that matches the localized surface plasmon (LSP) resonance, the NPG's temperature increases and transfers the heat through the Si$_3$N$_4$ and ITO layers to the VO$_2$ layer. By patterning not only the graphene sheet but also the Si$_3$N$_4$ and ITO layers we maximize the heat transfer to the VO$_2$ layer. Once the VO$_2$ layer's temperature increases above the phase transition temperature $T_c$, the VO$_2$ layer undergoes a transition from insulating to metallic phase.  The patterning of the VO$_2$ layer decreases its volume, thereby decreasing the heat required to drive the VO$_2$ layer over $T_c$, which in turn increases the sensitivity (NEP) of our proposed  NPG-VO$_2$ photodetector.  During this whole time a bias voltage $V_b$ is applied to the VO$_2$ layer, which upon phase transition triggers a current through the VO$_2$ layer, which can be detected electronically in about 1 ms, shorter than the detection time of typical VO$_2$ bolometers \cite{Chen2001}.

\begin{figure*}[t]
\begin{centering}
\includegraphics[width=18cm]{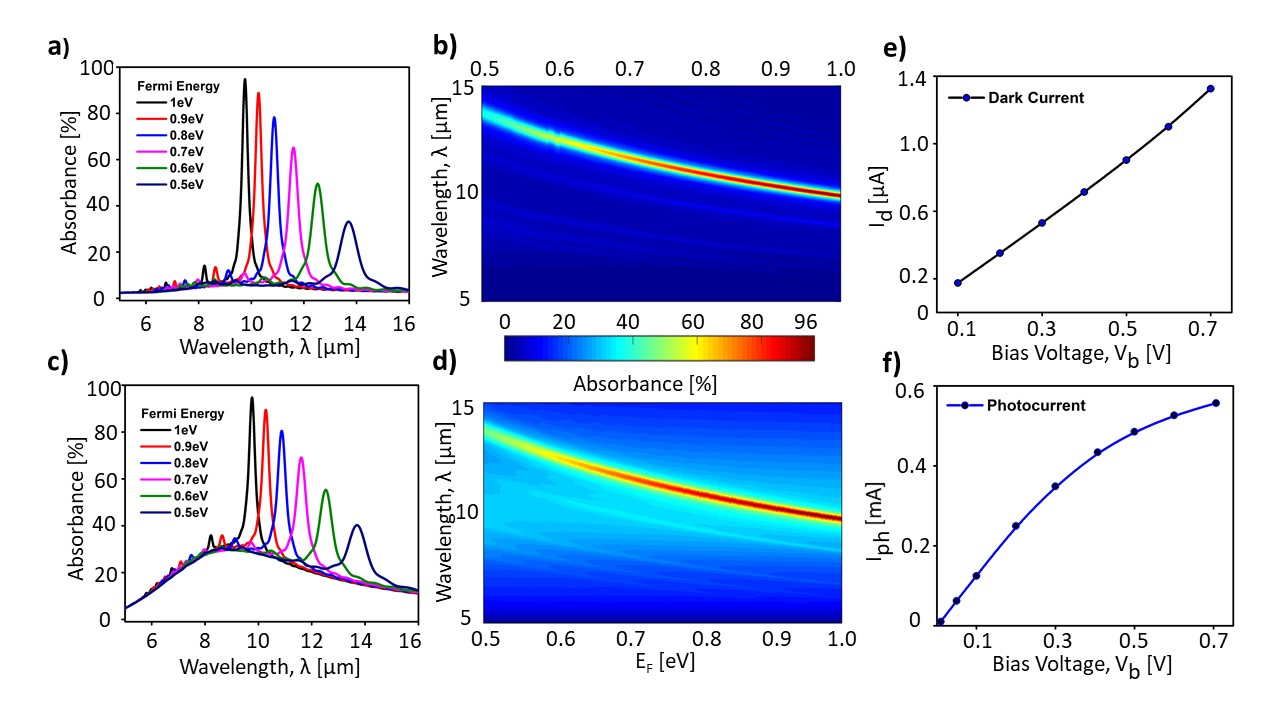}
\end{centering}
\caption{Absorbance and photocurrent of the NGP-VO$_2$ photodetector shown in Fig.~\ref{fig:graphene_VO2_detector} with nanopattern made of a hole diameter of 300 nm and a period of 450 nm. (a) Absorbance when VO$_2$ is in the insulating phase, calculated by means of the FDTD method. The resonance peaks of the LSP are clearly visible and tunable by means of a gate voltage that shifts the Fermi energy inside the NPG sheet.
Absorbance of mid-IR light in this wavelength regime in insulating VO$_2$ is very low.
(b) Absorbance as a function of wavelength and Fermi energy, showing the overall tunability of the LSP resonance peaks, when VO$_2$ is in the insulating phase.
(c) Absorbance of the NGP-VO$_2$ photodetector shown in Fig.~\ref{fig:graphene_VO2_detector} when VO$_2$ is in the metallic phase. The resonance peaks of the LSP are still clearly visible and tunable by means of a gate voltage that shifts the Fermi energy inside the NPG sheet. Compared with the absorbance for insulating VO$_2$ shown in (a), the metallic VO$_2$ layer absorbs mid-IR light over a large wavelength regime and exhibits a maximum of about 30\% at a wavelength of about $\lambda=9$ $\mu$m.
(d) Absorbance as a function of wavelength and Fermi energy, showing the overall tunability of the LSP resonance peaks, when VO$_2$ is in the metallic phase.
(e) $I_{\rm ph}-V_b$ characteristics of the NPG-VO$_2$ photodetector for insulating VO$_2$. Note that the scale of the current is given in $\mu$A and is thus very small compared to the photocurrent for metallic VO$_2$ shown in (f).
(f) $I_{\rm ph}-V_b$ characteristics of the NPG-VO$_2$ photodetector for metallic VO$_2$. Note that the scale of the current is given in mA and is thus very large compared to the photocurrent for insulating VO$_2$ shown in (e).
\label{fig:absorbance_current} }
\end{figure*}

 For the electronic response of the graphene sheet and the VO$_2$ layer to the incident mid-IR photons, we need to consider the intraband conductivity of graphene and the dielectric function of VO$_2$ in the insulating and metallic regimes.
 Using the linear dispersion relation, the intraband optical conductivity of graphene is \cite{Safaei2017,Paudel2017}
\be
\sigma _{\rm intra}(\omega ) = \frac{e^2}{\pi\hbar^2}\frac{2k_BT}{\tau^{-1} - i\omega}\ln \left[ 2\cosh \left( \frac{E_F}{2k_BT} \right) \right],
\ee
 which in the case of $E_F \gg {k_B}T$   is reduced to 
 \be
\sigma_{\rm intra}(\omega) = \frac{e^2}{\pi\hbar^2}\frac{E_F}{\tau^{-1} - i\omega }=\frac{2\varepsilon_m\omega_p^2}{\pi\hbar^2(\tau^{-1}-i\omega)},
\label{eq:sigma_intra}
 \ee
 where $\tau$ is determined by impurity scattering and electron-phonon interaction ${\tau ^{ - 1}} = \tau _{imp}^{ - 1} + \tau _{e - ph}^{ - 1}$ .
 Using the mobility $\mu$ of the NPG sheet, it can be presented in the form
 $\tau^{-1}=ev_F^2/(\mu E_F)$, where $v_F=10^6$ m/s is the Fermi velocity in graphene.
 $\omega_p=\sqrt{e^2E_F/2\varepsilon_m}$ is the bulk graphene plasma frequency.

 The dielectric function of VO$_2$ can be approximated 
 by means of a constant for the insulating phase \cite{Jepsen2006},
 \be
 \epsilon_i=\epsilon_\infty,
 \label{eq:eps_i}
 \ee
 and by means of the Drude formula for the metallic phase,
 \be
 \epsilon_m(\omega)=\epsilon_\infty-\frac{\Omega_p^2}{\omega(\omega+i\Gamma)}\left(1+\frac{ic\Gamma}{\omega+i\Gamma}\right),
 \label{eq:eps_m}
 \ee
 where $\Omega_p=\sqrt{Ne^2/\epsilon_0m^*}$ is the plasma frequency of VO$_2$, $\Gamma=e/m^*\mu_{\rm VO2}$ is the scattering rate in VO$_2$, with $\mu_{\rm VO2}=2$ cm$^2$/Vs being the mobility in VO$_2$, $N=1.3\times 10^{22}$ cm$^{-3}$ the free-carrier concentration in VO$_2$, and $m^*=2m_e$ the effective mass of the charge carriers in VO$_2$. $m_e$ is the free electron mass. $c$ is the fraction of the original velocity of the electron after scattering.

 We used the finite-difference time domain method (FDTD) to calculate the absorbance of the hybrid NPG-VO$_2$ photodetector as shown in Fig.~\ref{fig:absorbance_current} (a) when VO$_2$ is in the insulating phase and in Fig.~\ref{fig:absorbance_current} (c) when VO$_2$ is in the metallic phase.
 The resonance peaks due to the absorption of mid-IR light by localized surface plasmons (LSPs) in NPG are clearly visible and similar to the ones found in Refs. \cite{Safaei2017,SafaeiACS,Safaei2019,Shabbir2020}.
 The main difference is that the VO$_2$ layer exhibits very low absorbance (around 6\%) of mid-IR light in the insulating phase but a larger broadband absorption in the metallic phase.
 The thicker the metallic VO$_2$ layer is, the stronger is the absorption of mid-IR light. We chose a VO$_2$ layer thickness of 3 nm for two important reasons:
 Firstly, a thinner VO$_2$ layer results in weaker absorption and thus makes it easier for the photodetector to cool down after the incident mid-IR radiation is turned off.
 Secondly, the thinner VO$_2$ layer has a lower phase transition temperature of $T_c=52^\circ$C (=325 K) and thus requires less external heating for keeping the photodetector at an optimum operating temperature (see below).
 
 The absorbance resonance peak as a function of wavelength can be tuned by means of the Fermi energy of NPG, as shown in Fig.~\ref{fig:absorbance_current} (b) when VO$_2$ is in the insulating phase and in Fig.~\ref{fig:absorbance_current} (d) when VO$_2$ is in the metallic phase. Thus, the gate voltage allows the operating wavelength of the VO$_2$-NPG photodetector to be tuned in addition to the size and period of the nanopattern in graphene.

\begin{figure*}[tb]
\begin{centering}
\includegraphics[width=18cm]{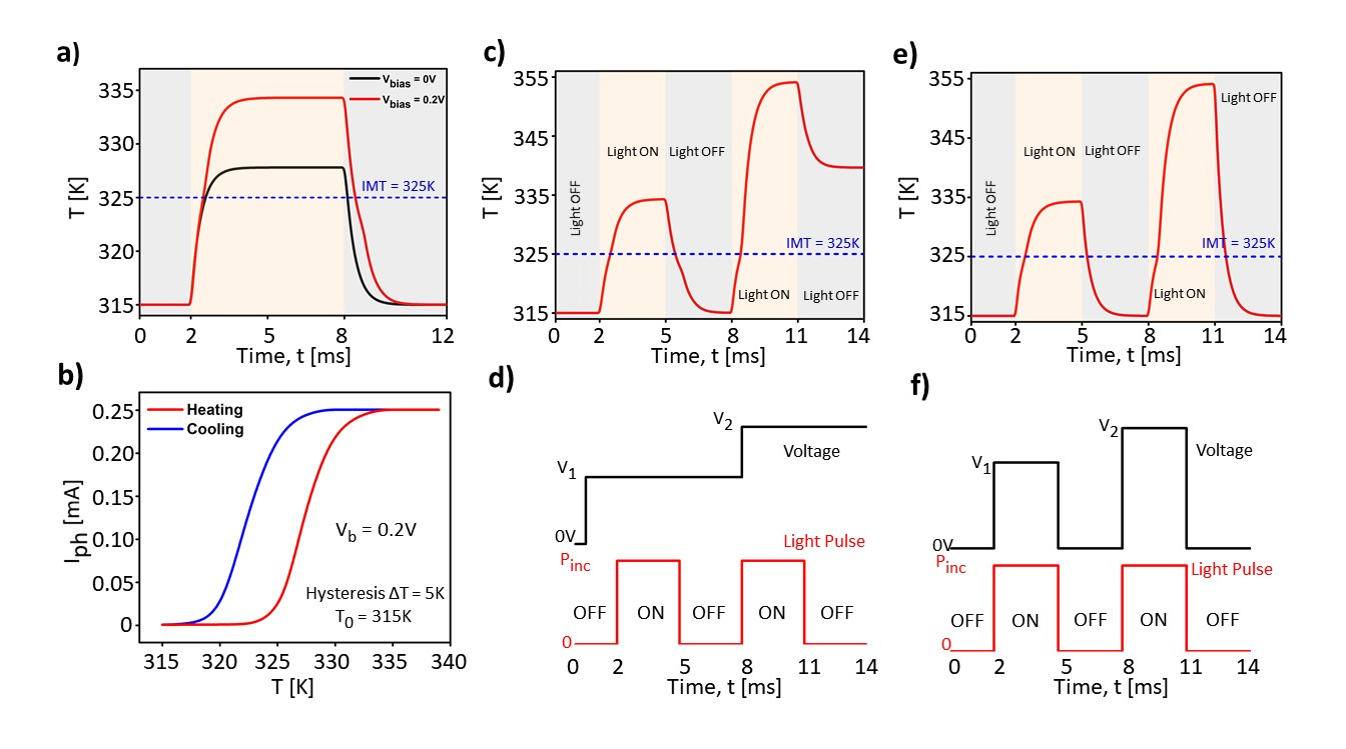}
\end{centering}
\caption{Photocurrent as functions of temperature and time of the NGP-VO$_2$ photodetector shown in Fig.~\ref{fig:graphene_VO2_detector} with nanopattern made of a hole diameter of 300 nm and a period of 450 nm. (a) Temperature $T$ of the VO$_2$ layer as a function of time, showing the photodetection process while the incident mid-IR light is turned on at time $t=2$ ms and subsequently turned off at time $t=8$ ms for constant applied bias voltages $V_b=0$ (black curve) and $V_b=0.2$ V (red curve).  (b) Photocurrent $I_{\rm ph}$ through the VO$_2$ layer as a function of Temperature $T$ for a constant applied bias voltage $V_b=0.2$ V during the heating (red curve) and cooling (blue curve) process. The base temperature of the substrate is kept at $T_0=315$ K. The temperature difference of the hysteresis is $\Delta T=5$ K.
(c) Temperature $T$ of the VO$_2$ layer as a function of time for two constant values of the bias voltages $V_b=0.2$ V and $V_b=0.25$ V.  (d) Bias voltage $V_b$ as a function of time $t$.
\label{fig:photodetection} }
\end{figure*}   
 
For modeling the operation of the proposed NPG-VO$_2$ mid-IR photodetector, we used COMSOL and the following theory for the thermoelectric properties of VO$_2$ close to the phase transition temperature $T_c$.
The VO$_2$ layer is operated around the IMT phase transition temperature $T_c$. The performance of the bolometric detection can be analyzed by means of the heat equation and a hysteresis model \cite{Almeida2004}. The heat equation reads 
\be
C\frac{{dT}}{{dt}} = \alpha P + {I^2}R(T) - G(T - {T_h}),
\label{eq:heat_capacity}
\ee
where $C$ is the heat capacity, $\alpha$ is the absorbance, $P$ is the power of the incident radiation, $I$ is the time-independent bias current, $R(T)$ is the temperature-dependent resistance, $G$ is the thermal conductivity of the heat sink, and $T_h$ is the time-independent temperature of the heat sink. The hysteretic behavior of $R(T)$ for VO$_2$ layer can be calculated by
 \be
R(T) = 17\exp \left( {\frac{{2553}}{{T + 273}}} \right)g(T) + 140,
\label{eq:resistance}
 \ee
where the semiconductor volume fraction is given by
\be
g(T) = \frac{1}{2} + \frac{1}{2}\tanh \beta \left[ {\delta \frac{w}{2} + {T_c} - \left( {T + {T_{pr}}P\left( {\frac{{T - {T_r}}}{{{T_{pr}}}}} \right)} \right)} \right],
\label{eq:volume_fraction}
 \ee
where $w$ is the width of the hysteresis, $\beta$ is a function of $dg/dT$ at $T_c$, $P(x)$ is an arbitrary monotonically decreasing function, and $\delta  = {\rm{sign}}\left( {dT/dt} \right)$.
The proximity temperature is given by
\be
{T_{pr}} = \delta \frac{w}{2} + {T_c} - \frac{1}{\beta }\arctan {\rm{h}}\left( {2{g_r} - 1} \right) - {T_r}.
\label{eq:Tpr}
\ee
Eqs. (\ref{eq:heat_capacity})-(\ref{eq:Tpr}) provide a simple method to describe the hysteretic behavior of the plasmonically driven bolometric photodetector. The incident power $P_{\rm inc}$ is then given by the energy pumped into the plasmonic nanostructure.
         
Using this thermoelectric theory and combining it with our FDTD results, we developed a photothermoelectric theory of the NPG-VO$_2$ heterostructure.
We present the current as a function of applied bias voltage $V_b$ (I-V characteristics) for insulating VO$_2$ in Fig~\ref{fig:absorbance_current} (e). and for metallic VO$_2$ in Fig.~\ref{fig:absorbance_current} (f). Note that the current flows only through VO$_2$, not through NPG. NPG is used only as a photothermoelectric heating element.
When the incident mid-IR light is off, a very weak dark current $I$ on the scale of $\mu$A is flowing when a bias voltage $V_b$ is applied on the order of 0.1 to 1.0 V.
In stark contrast, when the incident mid-IR light is on, a much larger light current $I$ on the scale of mA is flowing with the same applied bias voltage $V_b$. 
This effect is due to the phase transition of VO$_2$ between insulating and metallic phases.
We take advantage of this effect to develop the model of an ultrasensitive photodetector based on the NPG-VO$_2$ heterostructure.

After modeling the heating and cooling of the NPG-VO$_2$ heterostructure as a function of time, we identified the optimum photodetection process.
The resulting temperature $T$ of the VO$_2$ layer as a function of time $t$ and the photocurrent $I_{\rm ph}$ through the VO$_2$ layer as a function of temperature $T$ for a constant applied bias voltage $V_b=0.2$ V are shown in Fig.~\ref{fig:photodetection} (a) and (b).
We used the theoretical models and experimental values of conductivity, thermal conductivity, and heat capacity of VO$_2$ given in Refs.~\cite{Qazilbash2009,Zhong2011,Samanta2015,Ordonez2018}.
Our results reveal that the NPG-VO$_2$ photodetector has a response time on a time scale of the order of 1 ms, shorter than current microbolometers based on VO$_2$ alone \cite{Chen2001}.

The amount of Joule heating can be seen in Fig.~\ref{fig:photodetection}(a) as the difference in temperature between the cases $V_b=0$ V and $V_b=0.2$ V.
When the applied bias voltage $V_b$ is kept constant and is too large, the Joule heating prevents the photodetector from coolling down, as shown in Fig.~\ref{fig:photodetection} (c) and (d).
Such a behavior is problematic. In order to overcome this problem, one can apply a pulsed bias voltage $V_b(t)$, which ensures the cooling of the photodetector below the phase transition temperature $T_h$, as shown in Fig.~\ref{fig:photodetection} (e) and (f).

The responsivity $\cR$ of the NPG-VO$_2$ photodetector can be calculated by means of the formula \cite{Safaei2019}
\be
\cR=\frac{{({I_{\rm light}} - {I_{\rm dark}})R}}{{{P_{\rm inc}}}},
\ee
where $R$ is the resistance of VO$_2$ in the metallic phase, $I_{\rm light}$ is the light current when the incident light is on, $I_{\rm dark}$ is the dark current when the incident light is off, and $P_{\rm inc}$ is the power of the incident light.
The responsivity as a function of Fermi energy $E_F$ of NPG is shown in Fig.~\ref{fig:responsivity}.

\begin{figure}[htb]
\begin{centering}
\includegraphics[width=8cm]{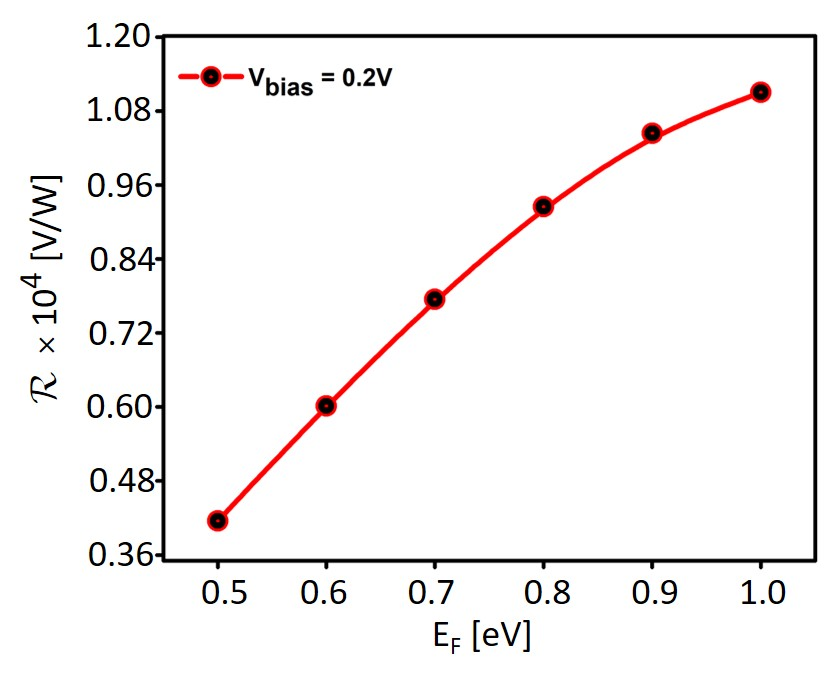}
\end{centering}
\caption{Responsivity of the NPG-VO$_2$ photodetector as a function of Fermi energy $E_F$ of NPG.
\label{fig:responsivity} }
\end{figure}

The sensitivity of the  NPG-VO$_2$ photodetector is determined by the noise-equivalent power NEP, being an important figure of merit for the performance of a photodetector. 
The NEP of a photodetector provides a measure for the minimum detectable power per 1 Hz of bandwidth.
The formula for NEP reads \cite{Safaei2019}
\be
{\rm NEP}=\frac{\nu_n}{\cR},
\ee
where
\be
\nu_n=\sqrt{\nu_t^2+\nu_b^2+\nu_f^2}
\ee
is the root-mean-square of the total noise voltage, which consists of the sum over all possible noise voltages, such as the thermal Johnson-Nyquist noise $\nu_t$, due to thermal motion of the charge carriers and independent of the bias voltage $V_b$, the shot noise $\nu_b$, due to the discrete nature of uncorrelated charge carriers, and the $1/f$ noise $\nu_f$, also called flicker noise, due to random resistance fluctuations.
The Johnson noise is given by \cite{Guo2020}
\be
\nu_t = \sqrt{4{k_B}TR},
\ee
where $k_B$ is the Botzmann constant, $T$ is the temperature, and $R$ is the resistance.
The shot noise is given by \cite{Guo2020}
\be
\nu_b = \sqrt{2eI_dR^2},
\ee
where $e$ is the elementary charge and $I_d$ is the dark current.
Since the dark current is very low and the NPG-VO$_2$ photodetector operates close to room temperature, the shot noise is much smaller than the Johnson noise.
Therefore, we can safely neglect the shot noise.
At a modulation frequency of $V_b$ of around 1 kHz we can also neglect the $1/f$ noise.
Using the NEP, we can calculate the detectivity of the NPG-VO$_2$ photodetector by means of the formula \cite{Safaei2019}
\be
D^*=\frac{\sqrt{A}}{\rm NEP},
\ee
where $A$ is the area of the photodetector.
The results of these figures of merit are shown in Table \ref{tab:merit}. 
By increasing the substrate temperature $T_0$, it is possible to decrease the hysteresis temperature difference $\Delta T$, which in turn results in an enhanced sensitivity NEP and a larger detectivity $D^*$. The details of these additional results are shown in the supplementary material.

\begin{figure}[htb]
\begin{centering}
\includegraphics[width=8cm]{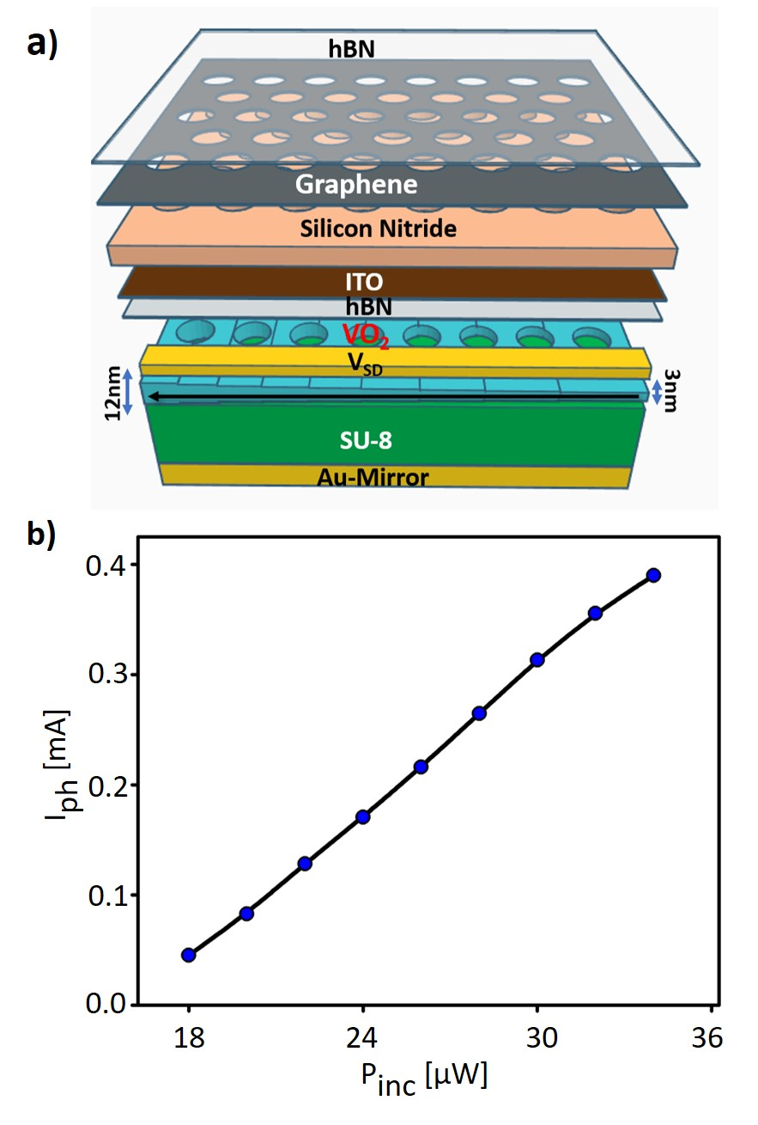}
\end{centering}
\caption{(a) Illustration of NPG-VO$_2$ photodetector with a VO$_2$ layer whose thickness has a gradient varying from 3 nm to 12 nm.
(b) Photocurrent $I_{\rm ph}$ as a function of input power $P_{\rm inc}$ of the mid-IR light for the photodetector shown in (a), which is linear due to the gradient in the thickness of the VO$_2$ layer.
\label{fig:photodetector_VO2gradient} }
\end{figure}

\begin{table}[h!]
  \begin{center}
    \begin{tabular}{||c|c|c|c|c|c||} % <-- Changed to S here.
    \hline\hline
      $\Delta T$ [K] & $P$ [$\mu$W] & $T_0$ [K] & $\cR [V/W]$ & NEP [fW/$\sqrt{Hz}$] & $D^*$ [Jones] \\
      \hline
      5.0 & 18.0 & 315.0 & 1.1$\times 10^4$ & 347 & 0.50$\times 10^{10}$ \\
      3.0 & 14.0 & 317.0 & 1.4$\times 10^4$ & 273 & 0.64$\times 10^{10}$ \\
      1.2 & 10.6 & 318.8 & 1.9$\times 10^4$ & 203 & 0.85$\times 10^{10}$ \\ \hline\hline
    \end{tabular}
    \caption{Figures of Merit of the NPG-VO$_2$ photodetector.}
     \label{tab:merit}
  \end{center}
\end{table}

Table \ref{tab:merit} shows that the detectivity $D^*$ of the NPG-VO$_2$ photodetector operating close to room temperature is close to $D^*$ of cryogenically cooled HgCdTe photodetectors. The NPG-VO$_2$ photodetector reaches a sensitivity close to VO$_2$ microbolometers while exhibiting a shorter detection time of around 1 ms and being able to detect photons also in the LWIR regime, which is impossible for VO$_2$ microbolometers.

Since we want to realize a linear dependence of the photocurrent $I_{\rm ph}$ as a function of input power $P_{\rm inc}$ of the mid-IR light, we add a gradient in the thickness of the VO$_2$ layer as shown in Fig.~\ref{fig:photodetector_VO2gradient} (a), i.e. the VO$_2$ layer thickness is varied from 3 nm to 12 nm.
The resulting linear photocurrent $I_{\rm ph}$ as a function of input power $P_{\rm inc}$ is shown in Fig.~\ref{fig:photodetector_VO2gradient} (b).
Indeed, the photocurrent $I_{\rm ph}$ is now linear as a function of the input power $P_{\rm inc}$, which provides an optimized mapping of $P_{\rm inc}$ onto $I_{\rm ph}$ for maximum dynamic range.

In conclusion, we present the model of an ultrasensitive mid-infrared (mid-IR) photodetector based on a heterostucture made of NPG and VO$_2$, thereby extending the responsivity of a VO$_2$ microbolometer to the LWIR domain. Moreover, this hybrid NPG-VO$_2$ photodetector has a narrowband absorption in the MWIR and LWIR that can be tuned by means of a gate voltage. Our results show that the NPG-VO$_2$ photodetector can reach a large responsivity $R\sim 10^4$ V/W, a detectivity $D^*\sim 10^{10}$ Jones, and a sensitivity in terms of NEP  $\mathrm{NEP}\sim 100$ fW/$\sqrt(\rm Hz)$ close to room temperature by taking advantage of the phase change of a thin VO$_2$ layer. The NPG sheet achieves an absorption of nearly 100\% due to localized surface plasmons (LSPs) around the patterned circular holes in a hexagonal lattice symmetry. The electrostatic gate potential can be used to tune the wavelength peak in the MWIR and LWIR regimes between 3 and 12 microns, thereby overcoming the intrinsic upper limit of 6 microns for microbolometers based on VO$_2$. Our COMSOL simulations show that the NPG-VO$_2$ photodetector is able to operate on a time scale of 1 ms, much shorter than the response times of current microbolometers based on VO$_2$ alone. Our proposed mid-IR photodetector reaches detectivities of cryogenically cooled HgCdTe photodetectors and sensitivities close to and field of view similar to VO$_2$ microbolometers while operating close to room temperature.

\begin{acknowledgments}
\textcolor{black}{We acknowledge support from NSF CISE-1514089.}
\end{acknowledgments}

\section{Supplementary Information}
In order to show that NPG is absolutely required to achieve mid-IR photodetection in the wavelength range of $\lambda=3$ $\mu$m to 12 $\mu$m as presented in this paper, we also calculated the absorbance and photocurrent for the cases (i) when graphene, Si$_3$N$_4$, and ITO are absent, (ii) when graphene is absent, and (iii) when graphene is not patterned. These results are compared with (iv) results for the NPG-VO$_2$ photodetector shown in Fig.~\ref{fig:graphene_VO2_detector}. All these results are shown in Table \ref{tab:comparison} in the case of $V_{\rm bias}=0.2$ V, base temperature $T_0=315$ K, incident radiation power $P_{\rm inc}=18$ $\mu$W, wavelength $\lambda=9.76$ $\mu$m, and VO$_2$ thickness of 3 nm.

\begin{figure}[htb]
\begin{centering}
\includegraphics[width=9.5cm]{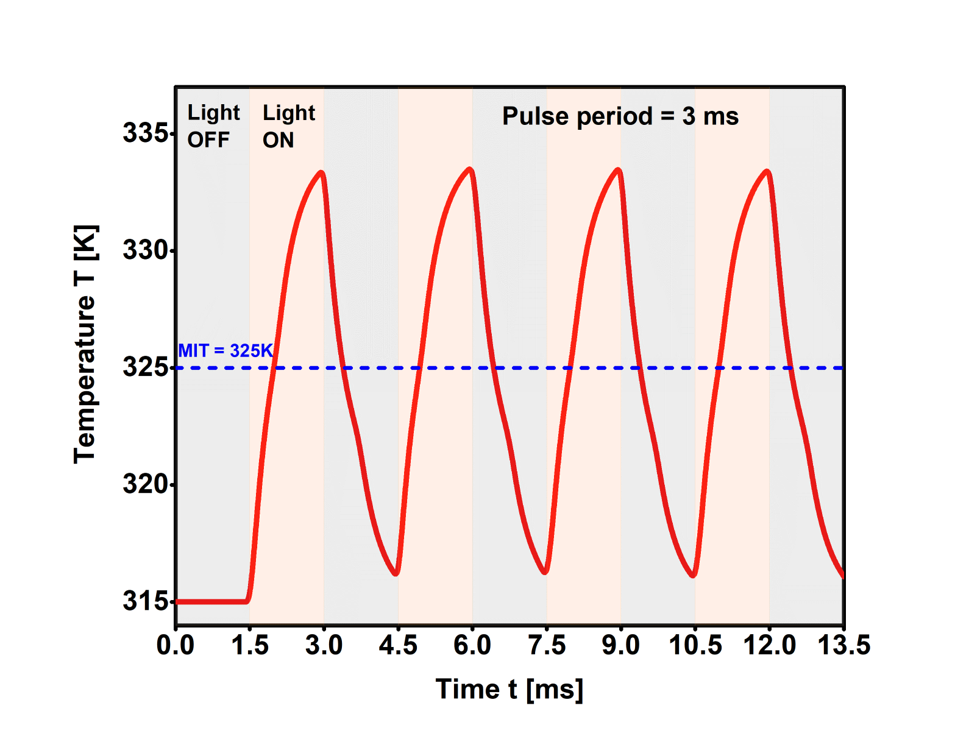}
\end{centering}
\caption{Temperature $T$ of the NPG-VO$_2$ photodetector as a function of time $t$ under pulsed mid-IR light illumination with period $\tau=3$ ms.
\label{fig:pulse} }
\end{figure}

\begin{table}[h!]
  \begin{center}
    \begin{tabular}{||c|c|c|c|c||} % <-- Changed to S here.
    \hline\hline
      Geometry & $T_{\rm max}$ [K] & $I_{\rm dark}$ [mA] & $I_{\rm light}$ [mA] & $I_{\rm ph}$ [mA] \\
      \hline
      (i) & 316.8 & 0.35$\times 10^{-3}$ & 0.369$\times 10^{-3}$ & 0.019$\times 10^{-3}$ \\
      (ii) & 317.4 & 0.35$\times 10^{-3}$ & 0.376$\times 10^{-3}$ & 0.026$\times 10^{-3}$ \\
      (iii) & 317.7 & 0.35$\times 10^{-3}$ & 0.381$\times 10^{-3}$ & 0.031$\times 10^{-3}$ \\ 
      (iv) & 334.3 & 0.35$\times 10^{-3}$ & 0.25 & 0.25 \\ \hline\hline
    \end{tabular}
    \caption{Comparison between geometries (i), (ii), (iii) without NPG and (iv) with NPG.}
     \label{tab:comparison}
  \end{center}
\end{table}

It can be seen from Table \ref{tab:comparison} that the cases (i) to (iii) achieve only a small photocurrent, in stark contrast to case (iv) that takes advantage of NPG. The reason for this large difference is the low absorbance of the structures (i) to (iii), as shown in Fig.~\ref{fig:absorbance_noNPG}.

\begin{figure}[htb]
\begin{centering}
\includegraphics[width=9.5cm]{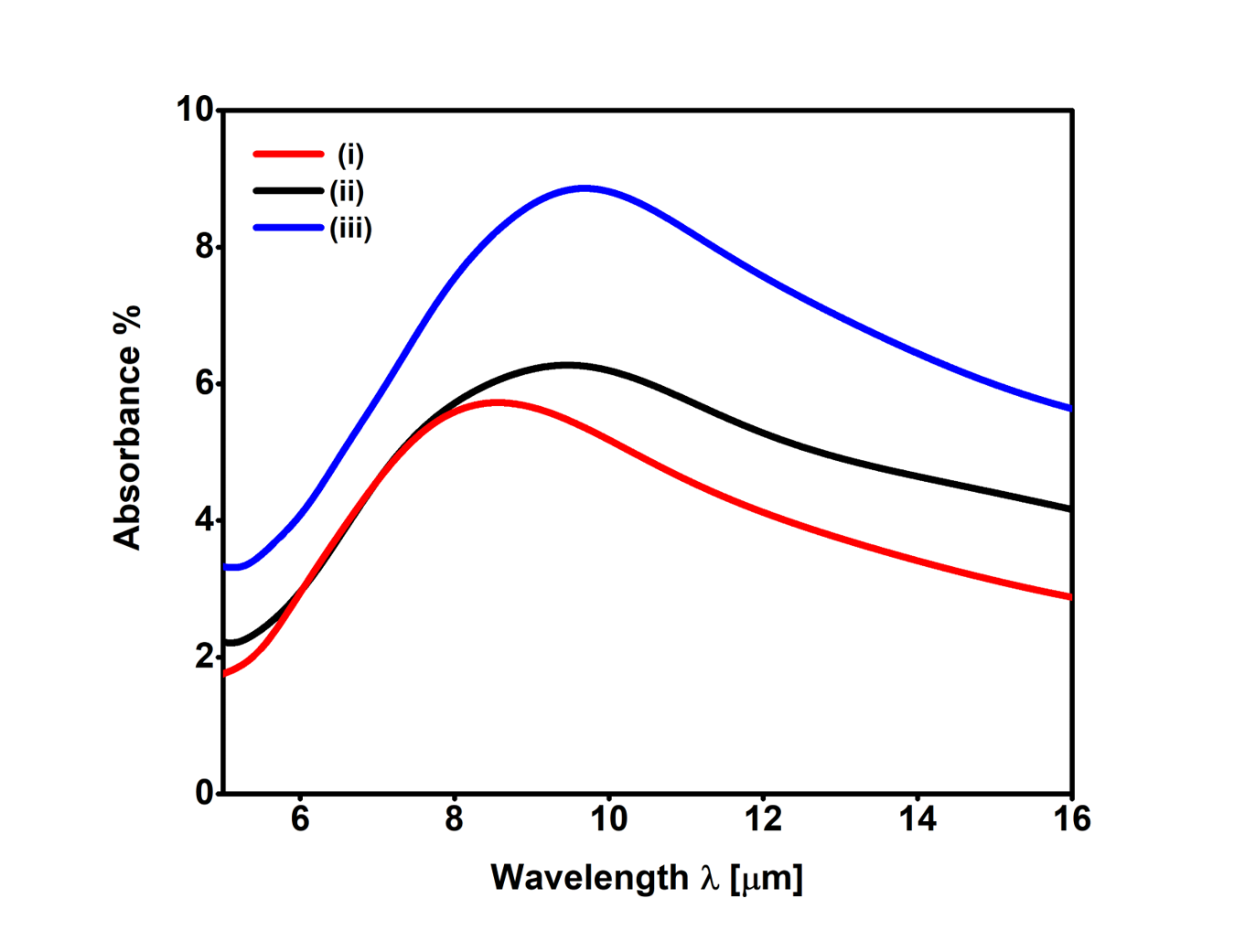}
\end{centering}
\caption{Absorbance as a function of wavelength $\lambda$ for the cases (i) when graphene, Si$_3$N$_4$, and ITO are absent, (ii) when graphene is absent, and (iii) when graphene is not patterned. These absorbances are much smaller than for case (iv) that takes advantage of NPG, as shown in Fig.~\ref{fig:absorbance_current}.
\label{fig:absorbance_noNPG} }
\end{figure}

In order to study the response of the NPG-VO$_2$ photodetector on a train of incident light pulses, we calculate the temperature $T$ as a function of time $t$ with pulse period $\tau=3$ ms. The result is shown in Fig.~\ref{fig:pulse}.
Thus, the response time of the NPG-VO$_2$ photodetector is of the order of 1 ms.

\begin{figure}[htb]
\begin{centering}
\includegraphics[width=9.5cm]{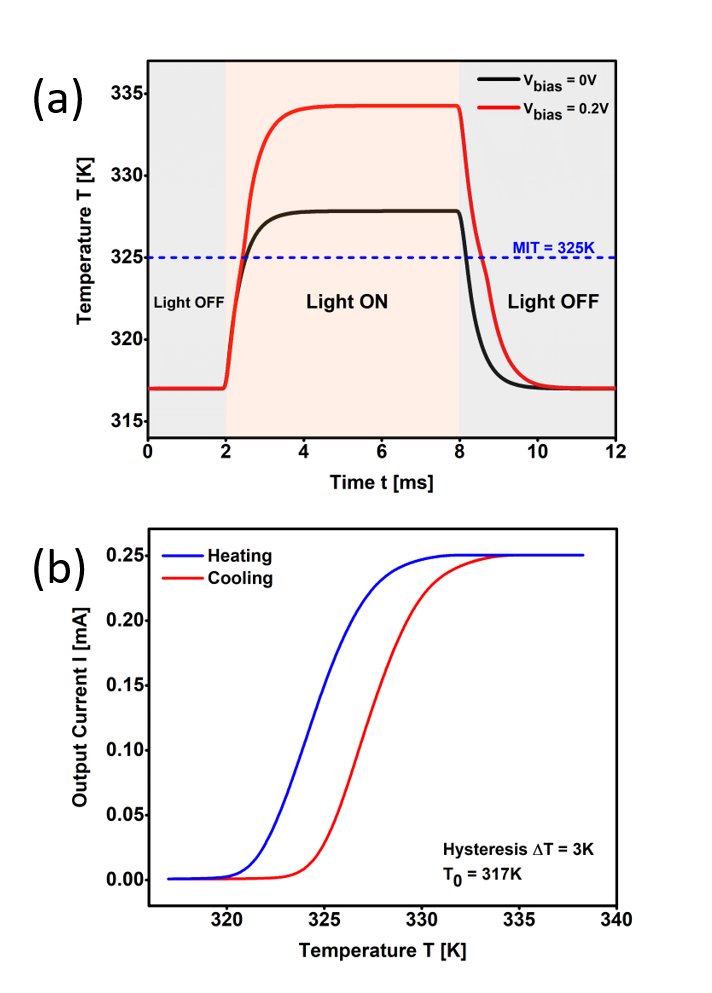}
\end{centering}
\caption{(a) Temperature $T$ of the VO$_2$ layer as a function of time, showing the photodetection process while the incident mid-IR light is turned on at time $t=2$ ms and subsequently turned off at time $t=8$ ms for constant applied bias voltages $V_b=0$ (black curve) and $V_b=0.2$ V (red curve).  (b) Photocurrent $I_{\rm ph}$ through the VO$_2$ layer as a function of Temperature $T$ for a constant applied bias voltage $V_b=0.2$ V during the heating (red curve) and cooling (blue curve) process. The base temperature of the substrate is kept at $T_0=317$ K. The temperature difference of the hysteresis is $\Delta T=3$ K.
\label{fig:photodetector_3K} }
\end{figure}

If the hysteresis curve of the photocurrent as a function of temperature can be made narrower with respect to the temperature $T$ \cite{Xu2012}, then it is possible to increase the base temperature of the substrate $T_0$. In that case, it is possible to reduce the required temperature difference $\Delta T$ for triggering the phase transition, which in turn allows us to also decrease the bias voltage $V_b$, thereby reducing the generated Joule heating power $P_J$ inside metallic VO$_2$.
The temperature $T$ of VO$_2$ as a function of time $t$ and the resulting photocurrent $I_{\rm ph}$ as a function of temperature for $T_0=317$ K and $\Delta T=3.0$ K is shown in Fig.~\ref{fig:photodetector_3K}, and  for $T_0=319$ K and $\Delta T=1.2$ K is shown in Fig.~\ref{fig:photodetector_1K}.

\begin{figure}[htb]
\begin{centering}
\includegraphics[width=9.5cm]{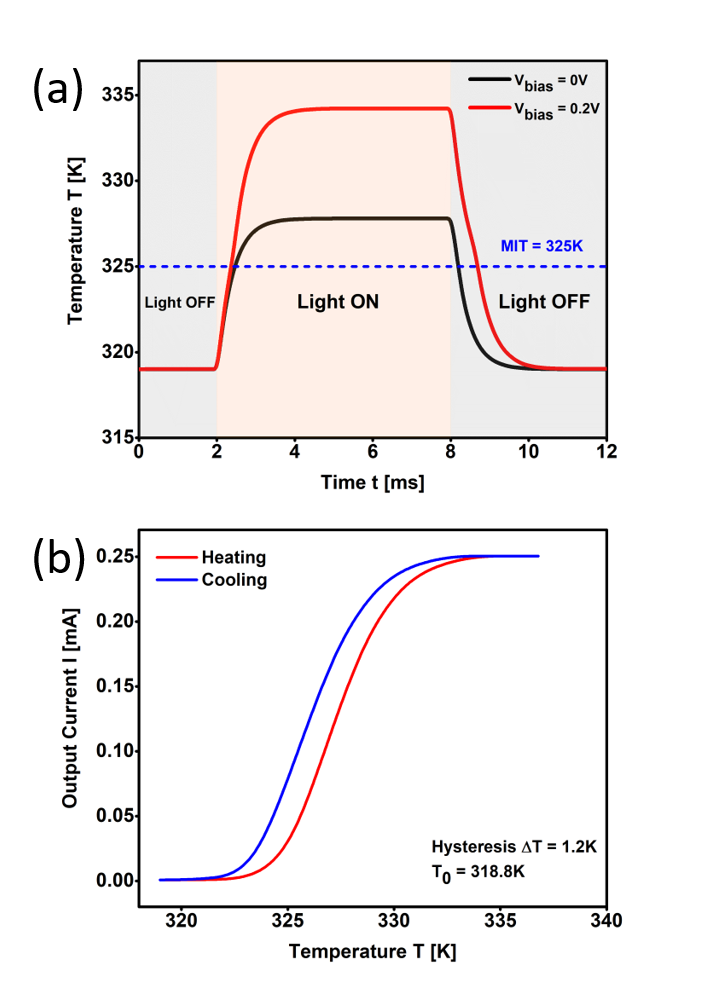}
\end{centering}
\caption{(a) Temperature $T$ of the VO$_2$ layer as a function of time, showing the photodetection process while the incident mid-IR light is turned on at time $t=2$ ms and subsequently turned off at time $t=8$ ms for constant applied bias voltages $V_b=0$ (black curve) and $V_b=0.2$ V (red curve).  (b) Photocurrent $I_{\rm ph}$ through the VO$_2$ layer as a function of Temperature $T$ for a constant applied bias voltage $V_b=0.2$ V during the heating (red curve) and cooling (blue curve) process. The base temperature of the substrate is kept at $T_0=319$ K. The temperature difference of the hysteresis is $\Delta T=3$ K.
\label{fig:photodetector_1K} }
\end{figure}

\end{document}